\documentclass{emulateapj}

\newcommand{\D}[1]{\, d #1 \,}
\newcommand{\vc}[1]{{\bf#1}}
\newcommand{\vch}[1]{\hat{\bf#1}}
\newcommand{\avg}[1]{\left< #1 \right>}
\newcommand{\ft}[1]{\widetilde{#1}}

\newcommand{\eps}{\varepsilon}

\newcommand{\hi}{H{\sc i}~}

\shorttitle{Velocity Spectrum for \hi}
\shortauthors{Chepurnov et al.}

\begin{document}

\title{Velocity Spectrum for \hi at High Latitudes}

\author{A. Chepurnov}
\author{A. Lazarian}
\author{S. Stanimirovi\'{c}}
\affil{Department of Astronomy, University of Wisconsin-Madison, 475 North Charter Street, Madison, WI 53706, USA}
\author{Carl Heiles}
\author{J. E. G. Peek}
\affil{Department of Astronomy, University of California, Berkeley, CA 94720}

\begin{abstract} 
In this paper we present the results of the statistical analysis of high-latitude \hi turbulence in the Milky Way. We have observed \hi in the 21 cm line, obtained with the Arecibo\footnote{The Arecibo Observatory is part of the National Astronomy and Ionosphere Center, which is operated by Cornell University under a cooperative agreement with the National Science Foundation} L-Band Feed Array (ALFA) receiver at the Arecibo radio telescope. For recovering of velocity statistics we have used the Velocity Coordinate Spectrum (VCS) technique. In our analysis we have used direct fitting of the VCS model, as its asymptotic regimes are questionable for Arecibo's resolution and given the restrictions from thermal smoothing of the turbulent line. We have obtained a velocity spectral index $3.87 \pm 0.11$, an injection scale of $140 \pm 80$ pc, and an \hi cold phase temperature of $52 \pm 11$ K. The spectral index is steeper than the Kolmogorov index and can be interpreted as being due to shock-dominated turbulence.
\end{abstract}


\section{Introduction}

Galactic interstellar medium (ISM) is turbulent over a very wide range of spatial scales (\citealt{Des00}, \citealt{Dick01}, \citealt{ES04}). This turbulence is a crucial parameter for understanding many astrophysical processes such as star formation, heat transfer, existence and evolution of ISM phases, cloud structure and dynamics, and cloud formation and destruction (see McKee \& Ostriker 2006, Lazarian et al. 2009).

Vivid signatures of interstellar turbulence include the ``Big Power Law'' of the electron density fluctuations (Armstrong et al. 1994), fractal structure of molecular clouds (Elmegreen \& Falgarone 1996, Stutzki et al. 1998), 
intensity fluctuations in channel maps (see Crovisier \& Dickey 1983, Green 1993, Stanimirovic et al. 1999, 
Deshpande, Dwarakanath \& Goss 2000, Elmegreen, Kim \& Staveley-Smith 2001). From the point of view of turbulence studies the velocity fluctuations reflected by the channel maps are of an evident importance. 

One of the main approaches for characterizing the ISM turbulence is based on using statistical descriptors. Many statistical tools for analyzing spectral-line data cubes have been attempted: the Principal Component Analysis (\citealt{HS97}), the Spectral Correlation Function (\citealt{Ros99}) and velocity centroids method (\citealt{EL05}, Ossenkopf et al. 2006, Esquivel et al. 2007). Wavelets, in particular $\Delta$-variance, have been shown to be very useful in studying inhomogeneous data (see Ossenkopf, Krips \& Stutzki 2008ab). These tools can be employed to investigate intensity fluctuations in spectral-line data cubes that carry information on velocity turbulence. Nevertheless, the direct relation between the underlying statistics of the velocity and the measures available with the techniques above is far from being straightforward (see a discussion in Lazarian 2009). 

The most direct and straightforward way of dealing with velocity fluctuations is to analyze the statistics of the Doppler-shifted spectral lines, most fully represented by the statistics of the position-position-velocity or PPV data cubes. However, relating the fluctuations of intensity in the PPV domain with the underlying 3-D velocity and density statistics is a problem that has been only recently addressed (see Lazarian \& Pogosyan 2000, Lazarian 2009). Lazarian \& Pogosyan (2000) developed the Velocity-Channel Analysis (VCA) technique which connected observed intensity fluctuations with the underlying density and velocity fluctuations by manipulating velocity resolution of
PPV data cubes.

In this paper we apply a new statistical technique, the Velocity Coordinate Spectrum (VCS) proposed initially in \cite{LP00} and elaborated in \cite{LP06} and \cite{CL06} (hereafter LP06 and CL06, respectively), on high-latitude \hi observations obtained with the Arecibo radio telescope as a part of the all-sky survey undertaken by GALFA-\hi. GALFA-\hi is a consortium for Galactic studies with the Arecibo L-band Feed Array (ALFA). The survey specifications and strategy are described in Stanimirovic et al. (2006), and data reduction methods are described in Peek and Heiles (2008). GALFA-\hi datasets have high spatial and velocity dynamic range and offer a good opportunity for testing the VCS technique. Earlier this technique was tested using synthetic observations \citep{CL09}.  

The structure of this paper is organized as follows. In Sect. \ref{sect:hi} we describe briefly the \hi data used in this paper. In Sect. \ref{sect:vcs} we review the VCS technique and address several important issues that need to be considered before applying VCS to \hi data. In Sect. \ref{sect:app} we apply the VCS technique on the HI data. The discussion of our results, advantages and limitations of VCS are provided in Sect. \ref{sect:dis}. Applicability of asymptotic studies to our data is discussed in Appendix.

\section{HI data} \label{sect:hi}

We observed a high-latitude Galactic region,  $16 \times 7$ square degrees large and centered on $\alpha={\rm 2^{h}15^{m}}$, $\delta={\rm +9^\circ30^m}$,  in May and June of 2005 using Arecibo's ALFA receiver and the GALSPECT spectrometer. ALFA is a 7-element focal-plane array primarily designed for 21-cm observations. GALSPECT is a special-purpose spectrometer for Galactic science with ALFA. GALSPECT has a spectral resolution of 0.18 km s$^{-1}$, and a fixed bandwidth of 1380 km s$^{-1}$. Each of the 7 beams of ALFA has a 3.35 arcminute FWHM beam width with a beam ellipticity of 0.2. The region contains much low-velocity, high-latitude Galactic HI, as well as a sub-complex of Very-High Velocity Clouds whose analysis is detailed in \cite{Peek2007}. The region was observed in a `basket-weave' or meridian-nodding mode, interlacing scans from day to day.

Data were reduced using the standard GALFA-HI reduction strategy  \citep{Peek2007}, without the first sidelobe correction. The final data cube was at the end scaled to the equivalent region in the Leiden-Dwingeloo Survey (LDS; \cite{Hartmann1997}) for a single, overall gain calibration. 

ALFA's  pixels are known to have asymmetric first sidelobes, as well as significant stray radiation, i.e. unmapped, distant sidelobes. These can contaminate the data slightly -- between 50\% and 70\% of the flux is in the main beam, 10\% to 20\% is in the first sidelobe and 20\% to 30\% is in stray radiation, depending upon which ALFA pixel is measured\footnote{see C. Heiles, 2004, \\ \texttt{www2.naic.edu$/$alfa$/$memos$/$alfa$\_$bm2.pdf}}. Unpublished work by Carl Heiles and Tom Troland leads us to believe that the stray radiation does not come from large angular distances from the main beam. This information is corroborated by the fact that the LDS spectra are quite consistent with our observed spectra, once scaled - if much of the flux came from sidelobes more distant than $36'$, the spectra would look significantly dissimilar.

\begin{figure}
\begin{center}
\plotone{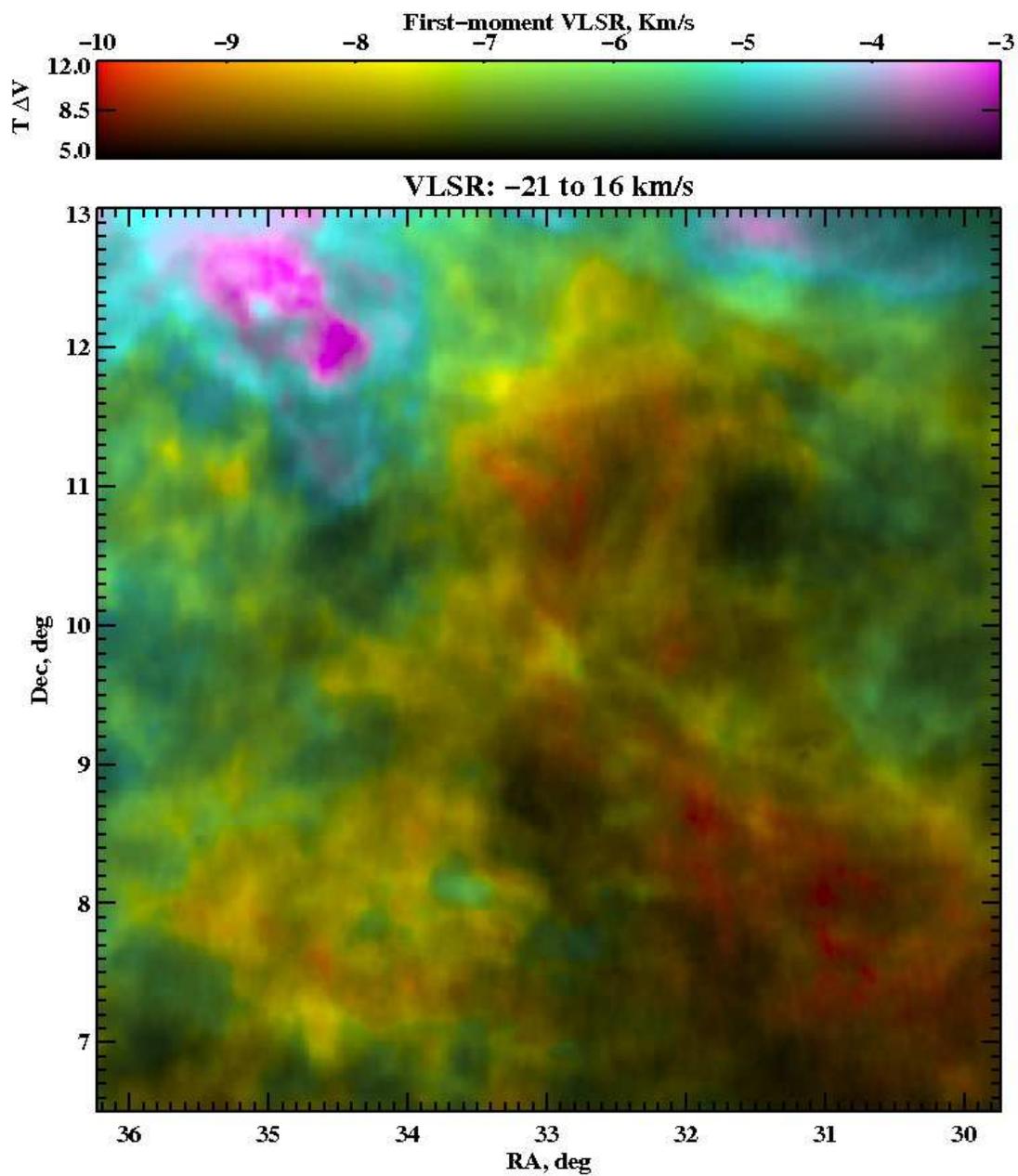}
\end{center}
\caption{The intensity-velocity map of the analyzed region. In this image, color indicates the first-moment velocity and brightness the integrated line intensity over the VLSR range -21 to 16 km/s. \label{fig:data}} 
\end{figure}

\begin{figure}
\begin{center} 
\plotone{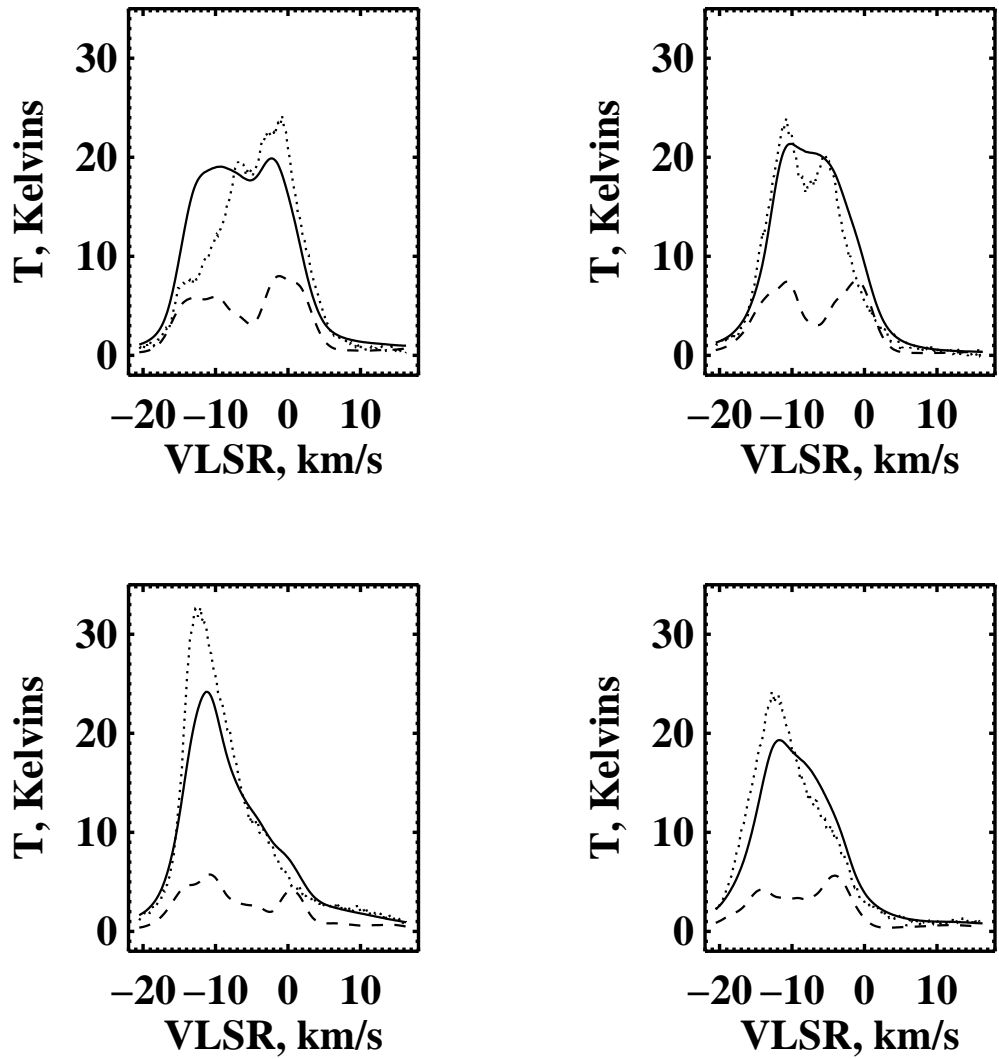}
\end{center}
\caption{\hi line profiles within the analyzed region. Spectra for the four quadrants of the image in Figure \ref{fig:data}. The spatial arrangement of the panels is the same as for the image quadrants. For each panel, the solid profile is the average of all spectra in the quadrant, dashed profile is the r.m.s. temperature over the quadrant and the dotted profile is the spectrum of the center of the quadrant.\label{fig:prof}} 
\end{figure}

In what follows we use approximately homogeneous $6^\circ.5 \times 6^\circ.5$ region, shown on Fig. \ref{fig:data}. The \hi profiles throughout the analyzed region are presented on Fig. \ref{fig:prof}. The average \hi spectra derived from four image quadrants are relatively similar suggesting that we can treat the whole region as being relatively homogeneous.

\section{The VCS technique} \label{sect:vcs}

\subsection{Basic Introduction}

The VCS technique is based on calculating the 1-D power spectrum of intensity fluctuations along the velocity axis, $P_1(k_v)$. This spectrum varies with the angular resolution of the telescope as it is illustrated in Figure~3. By investigating how $P_1$  changes when changing from high to low angular resolution, we can estimate the power spectrum of velocity fluctuations and ISM parameters such as temperature and Mach number.

\subsection{Theoretical Considerations}

We briefly overview the main analytical results from LP00, LP06 and CL06 which are relevant for this paper\footnote{Please note that we do not deal here with the VCS studies of turbulent volumes where self-absorption is important, or with saturated absorption lines \citet{LP08}}. Before going into derivations, we first state our main assumptions.

We assume that the observed HI intensity fluctuations arise from turbulence which can be characterized by two power spectra: the power spectrum of velocity\footnote{Much unfortunate confusion in the literature stems from the fact that the spectral indexes of turbulence may differ by a factor of 2 depending whether the integration over two $k$ directions is performed or not. For instance, the frequently quoted number for the Kolmogorov spectral index is $\alpha_v=5/3$, which is obtained after the aforementioned integration. We deal with power spectra that are not integrated over $k^2 dk$, thus the Kolmogorov spectrum index in this work is $\alpha_v=11/3$. This definition of the spectral index is consistent with our earlier papers.} $P_v\sim k^{-\alpha_v}$, and the power spectrum of emissivity (proportional to density in the case of \hi observations), $P_{\eps}\sim k^{-\alpha_{\eps}}$. Here $k$ is the wavevector in the ordinary 3D space (i.e. $k\sim 1/l$, where $l$ is a spatial scale). Power spectra $P_v$ and $P_{\eps}$ determine energy distribution of turbulent motions and density fluctuations in space. Both $P_v$ and $P_{\eps}$ contribute to the distribution of intensity fluctuations in the PPV space. Power spectra of velocity and density are Fourier transforms of the correlation functions of velocity and density, respectively. Those, however, are not directly available from observations. Thus the approach first adopted in LP00 was to study proper PPV statistics and relate them to the underlying structure function of velocity and correlation function of emissivity. 

We start with the expression for a spectral line signal, or an \hi intensity measured at velocity $v_0$ and at a given beam position $\vch{e}$: 
\begin{equation} \label{eq:sig_rat}
S(\vch{e},v_0)
  = \int w(\vch{e},\vc{r}) \D{\vc{r}} \eps(\vc{r}) f(v_r(\vc{r}) + v_r^{reg}(\vc{r}) - v_0),
\end{equation}
where $\eps$ is normalized emissivity, $v_r^{reg}$ is a line-of-sight component of the regular velocity (e.g. the velocity arising from the galactic velocity shift), $v_r$ is the line-of-sight component of the random turbulent velocity, $f$ denotes the convolution between the spectrometer channel sensitivity function\footnote{I.e. amplitude-frequency characteristic of a spectrometer channel normalized to its integral value with frequency in velocity units} and the Maxwellian distribution of velocities of gas particles, defined by temperature of emitting medium. The window function $w$ is defined as follows: 
\begin{equation} \label{eq:w}
w(\vch{e},\vc{r})
  \equiv \frac{1}{r^2} w_b(\vch{e},\vch{r}) w_\eps(\vc{r}),
\end{equation}
where $w_b$ is an instrument beam, pointed at the direction $\vch{e}$, which depends on angular coordinates $\vch{r}$, while $w_\eps$ is a window function defining the extent of the observed object.

The Fourier transform of a spectral line\footnote{Variable $k_v$ plays here the role of $k_z$ in LP00, being, however, different in dimension ($k_z = b k_v$, see Eq. (\ref{eq:uz_lin})). We use it here to avoid complications when $b=0$.} can be expressed as:
\begin{equation} \label{eq:sig_ft}
\begin{array}{ll}
\ft{S}(\vch{e},k_v) 
  & \equiv \frac{1}{2\pi} \int_{-\infty}^\infty S(v_0) e^{-i k_v v_0} \D{v_0} \\
  & = \ft{f}(k_v) \int w(\vch{e},\vc{r}) \D{\vc{r}} \cdot \\
  &   \eps(\vc{r}) \exp(-i k_v (v_r(\vc{r}) + v_r^{reg}(\vc{r}))).
\end{array}
\end{equation}
$\ft{S}(\vch{e},k_v)$ is a function of the direction of observation determined by the vector $\vch{e}$ and can be easily calculated from an observed PPV data cube. $k_v$ is the wavevector in the velocity space and $k_v\sim 1/v$. 

If we correlate $\ft{S}$ taken in two directions, pointed by $\vch{e}_1$ and $\vch{e}_2$, we get the following measure, which can be used as a starting point for the mathematical formulation of the VCS technique, as well as the VCA technique:
\begin{equation} \label{eq:K12_def} 
\begin{array}{ll}
K(\vch{e}_1,\vch{e}_2,k_v) 
  & \equiv \avg{\ft{S}(\vch{e}_1,k_v) \ft{S}^*(\vch{e}_2,k_v)} \\ 
  & =\ft{f}^2(k_v) \int w(\vch{e}_1,\vc{r}) \D{\vc{r}} \int w(\vch{e}_2,\vc{r'}) \D{\vc{r'}} \cdot\\
  & \avg{\eps(\vc{r})\eps(\vc{r'})} \avg{\exp(-i k_v (v_r(\vc{r})-v_{r'}(\vc{r'})))}  \cdot\\
  & \exp(-i k_v (v_r^{reg}(\vc{r})-v_{r'}^{reg}(\vc{r'}))),
\end{array}
\end{equation}
where $\langle...\rangle$ denotes averaging\footnote{Formally, this is an ensemble averaging, which is a mathematically rigorous concept, while our case of galactic \hi study, spatial averaging is applicable. This can be done by averaging over pairs of $\vch{e}_1,\vch{e}_2$ while keeping the distance between $\vch{e}_1,\vch{e}_2$ constant.} and where we assumed that gas velocity and emissivity are uncorrelated. The latter is not a necessary condition as LP00 showed that important regimes of the statistical study can be recovered even if the two quantities are correlated to a maximal degree. In addition, studies of synthetic maps obtained using 3D MHD simulations that exhibit velocity and density correlations (see Lazarian et al. 2000) demonstrate that this assumption does not significantly affect the final result. 

The first averaging in the last equation gives us the emissivity correlation function $C_\eps(\vc{r}-\vc{r'})$, which for \hi translates into the correlation function of overdensity, i.e. $\avg{\rho}^2+C_{\Delta\rho}$, where $C_{\Delta\rho}$ is a correlation function of density fluctuations. An average of the exponent can be performed under the assumption that the velocity statistics are Gaussian\footnote{We assume that the velocity field has a Gaussian Probability Distribution Function (PDF). The latter is fulfilled to high accuracy in both experimental (Monin \& Yaglom 1976) and numerical (Biskamp 2003) data.} (see LP00):
\begin{equation} 
\begin{array}{ll}
&\avg{\exp(-i k_v (v_r(\vc{r})-v_{r'}(\vc{r'})))} \\
&  = \exp\left(-\frac{k_v^2}{2} \avg{(v_r(\vc{r})-v_{r'}(\vc{r'}))^2}\right).
\end{array}
\end{equation}
To proceed, we assume that the beam separation and the beam width are both small enough that we can neglect the difference between $v_r$ and $v_z$ (we consider $z$-axis to be a bisector of the angle between beams). We also assume that $v_z^{reg}(\vc{r})$ depends only on $z$ and admits a linear approximation:
\begin{equation} \label{eq:uz_lin} 
v_z^{reg}(z)
  = b(z-z_0)+v_{z,0}^{reg},
\end{equation}
where $b$ characterizes regular velocity shear. The case in which velocity shear $b$ arises from Galaxy rotation is discussed in detail in LP00.

If we introduce a velocity structure tensor projection:
\begin{equation}
D_{vz}(\vc{r}-\vc{r'})
  \equiv \avg{(v_z(\vc{r})-v_z(\vc{r'}))^2}, 
\end{equation}
then Eq.~(\ref{eq:K12_def}) can rewritten as:
\begin{equation} \label{eq:K12_inter} 
\begin{array}{ll}
K(\vch{e}_1,\vch{e}_2,k_v)
  & =\ft{f}^2(k_v) \cdot \\
  & \int w(\vch{e}_1,\vc{r}) \D{\vc{r}} \int w(\vch{e}_2,\vc{r'}) \D{\vc{r'}} C_\eps(\vc{r}-\vc{r'}) \cdot \\
  & \exp \left(-\frac{k_v^2}{2} D_{vz}(\vc{r}-\vc{r'}) - ik_v b(z-z') \right). 
\end{array}
\end{equation}

Further transformations lead to: 
\begin{equation} \label{eq:K12} 
\begin{array}{ll}
K(\vch{e}_1,\vch{e}_2,k_v) 
  & =\ft{f}^2(k_v) \int g(\vch{e}_1,\vch{e}_2,\vc{r}) \D{\vc{r}} \cdot\\
  & C_\eps(\vc{r}) \exp \left(-\frac{k_v^2}{2} D_{vz}(\vc{r}) - ik_v b z \right), 
\end{array}
\end{equation}
where geometric factor $g$ is given by
\begin{equation}
g(\vch{e}_1,\vch{e}_2,\vc{r}) \label{eq:g_orig}
  \equiv \int w(\vch{e}_1,\vc{r'})w(\vch{e}_2,\vc{r'}+\vc{r}) \D{\vc{r'}}, 
\end{equation} 
and $\ft{f}$ is a Fourier transform of the effective channel sensitivity function $f$.

The measure $K$ given by Eq.~(\ref{eq:K12}) depends both on the velocity wavevector $k_v$ as well as on the angular distance between the vectors $\vch{e}_1$ and $\vch{e}_2$. If we integrate over $k_v$ within an interval defined by the required channel width, we arrive to the formalism of the VCA technique. In the opposite limit, if we take coincident vectors $\vch{e}_1$ and $\vch{e}_2$ to obtain a one-dimensional spectrum $P_1$, we arrive at the starting measure for the VCS technique:
\begin{equation} \label{eq:P1_orig} 
\begin{array}{ll}
P_1(k_v)
  & \equiv K(\vch{e}_1,\vch{e}_2,k_v)\vert_{\vch{e}_1=\vch{e}_2} 
  = \ft{f}^2(k_v) \int g(\vc{r}) \D{\vc{r}} \cdot \\
  & C_\eps(\vc{r}) \exp \left(-\frac{k_v^2}{2} D_{vz}(\vc{r}) - i k_v b z \right).
\end{array}
\end{equation}
where $g(\vc{r}) \equiv g(\vch{e},\vch{e},\vc{r})$ and $\vch{e}$ is a beam direction.

As it was shown in LP00 and CL06, Eq. (\ref{eq:P1_orig}) has two asymptotic spectral regimes which depend on beamwidth. For the ``high resolution mode'' ($k_v$ is less than unity over r.m.s. velocity on the beam scale) the slope of $P_1$ is $2/(\alpha_v - 3)$, otherwise, in the ``low resolution mode'' it is $6/(\alpha_v - 3)$. We have assumed here a steep density spectrum\footnote{Whether or not the last claim is true can be established with the analysis of column density maps. As we neglect the effects of self-absorption, the column densities can be obtained via v-integration of PPV cubes.  Naturally, in the column density maps the spectrum is affected only by density and its slope is $\alpha_\eps$. The situation is a bit more complicated when the density is shallow ($\alpha_\eps<3$), which is the case in high Mach number turbulence (see Beresnyak, Lazarian \& Cho 2006) and the density combines with velocity to affect the $P_1$ slope. For a more detailed analysis, see LP00. To measure the slope of the density spectrum the column density image can be used (see Section \ref{sect:p1mod}).}  (i.e. when $\alpha_\eps>3$). The applicability of asymptotics depends on many factors and usually requires direct calculation of $P_1$. In our analysis we did direct calculations of Eq. (\ref{eq:P1_orig}) as the asymptotic regime assumption is questionable for the analyzed PPV cube (see Appendix for more details).




We further include the possibility that the observer is located inside or close to the emitting structure (i.e. lines of sight are converging as illustrated in Fig. \ref{fig:geo}). This affects the geometric factor $g$, defined by Eqs. (\ref{eq:g_orig}) and (\ref{eq:w}). CL06 showed that for a Gaussian beam with the radius $\theta_0$, the correspondent $g(\vc{r})$ for converging lines of sight can be reduced to:
\begin{equation} \label{eq:wa_clos}
\begin{array}{ll} 
g(\vc{r})  
  = & \frac{1}{\pi \theta_0^2} \int_0^\infty \frac{
    w_\eps(z') w_\eps(z'+|z|)}{z'^2+(z'+|z|)^2}  \\ 
  & \cdot \exp \left(-\frac{R^2}{\theta_0^2 (z'^2+(z'+|z|)^2)}\right)\D{z'},
\end{array}
\end{equation}
\noindent where $\vc{R} \equiv (x,y)$. If we set $w_\eps$ as follows:
\begin{equation} \label{eq:weps_tophat} 
  w_\eps(z) 
    = \left\{
      \begin{array}{ll}
        1, & z \in [z_0,z_1] \\
        0, & z \notin [z_0,z_1]
      \end{array}
    \right.
\end{equation}
\noindent where $z_0$ and $z_1$ are inner and outer borders of an emitting layer in a given direction, we have the following expression for $g(\vc{r})$:
\begin{equation} \label{eq:wa_clos_appr}
\begin{array}{ll}
g(\vc{r}) 
  & \approx \frac{-1}{2\sqrt{\pi} \theta_0 R z} \cdot
    \frac{
      \arctan\left(1+\frac{2 z_0}{z}\right)+\arctan\left(1-\frac{2 z_1}{z}\right)
    }{
      \left(2 z_0^2 + p z^2\right)^{-\frac{1}{2}} - 
      \left(2 z_1^2 - p z^2\right)^{-\frac{1}{2}} 
    } \cdot\\
  & \left( 
      \mathrm{erf}\left(\frac{R}{\theta_0 \sqrt{2 z_0^2 + p z^2}} \right) - 
      \mathrm{erf}\left(\frac{R}{\theta_0 \sqrt{2 z_1^2 - p z^2}} \right)
    \right),
\end{array}
\end{equation}
\noindent where
\begin{equation} 
  p
    = \frac{z_1+z_0}{z_1-z_0}.
\end{equation}

We discuss in Section \ref{sect:vcs_appl} our selection of $z_0$ and $z_1$ limits.

We now need to express $D_{vz}$ through the velocity spectrum. If we assume that $\vc{v}(\vc{r})$  is solenoidal\footnote{Numerical simulations show that most of the energy resides in solenoidal motions even for compressible driving.} with power-law power spectrum having cutoff at large scales, the velocity power spectrum can be written in as follows (see, for example, \citealt{Les91}):
\begin{equation} \label{eq:Fu} 
F_{ij}(\vc{k})
  = \frac{V_0^2}{k^{\alpha_v}} e^{-\frac{k_0^2}{k^2}} \left(\delta_{ij} - \frac{k_i k_j}{k^2}\right),
\end{equation}
\noindent where $V_0^2$ is the velocity power spectrum amplitude, $\alpha_v$ is the velocity spectral index, $k_0=2\pi/L_v$ is the cutoff wavevector bound with the injection scale $L_v$, and  $i$ and $j$ are the component indexes. Then $D_{vz}$ can be represented as
\begin{equation} \label{eq:Dz_Fu} 
D_{vz}(\vc{r})
  = 2 \int \D{\vc{k}} (1 - e^{i\vc{k}\vc{r}}) \hat{z}_i\hat{z}_j F_{ij}(\vc{k}).
\end{equation} 
\noindent(Summation over repeating indexes is assumed here.)

\section{Data Analysis} \label{sect:app}

\subsection{Model of $P_1$} \label{sect:p1mod}
We first calculate the 2D spatial power spectrum of the \hi column density image as this directly provides us with $\alpha_\eps$. This spectrum is shown in Figure \ref{fig:rho} and has a power-law slope of $\sim3$ (the significance of this slope is discussed  in Section \ref{sect:dis}). This simplifies our analysis as the density (emissivity) correlation function ($C_{\epsilon}$) has in this case weak (logarithmic) dependence on $\vc{r}$ and can be factored out of the integrand in Eq. (\ref{eq:P1_orig}). This results in further expression for $P_1$:
\begin{equation} \label{eq:P1_mod}
P_{1,mod}(k_v)
  = \ft{f}^2(k_v) P_0 \int g(\vc{r}) \D{\vc{r}} \exp \left(-\frac{k_v^2}{2} D_{vz}(\vc{r})\right) + N_0,
\end{equation}
\begin{equation} \label{eq:FT_f}
  \ft{f}(k_v) = \frac{1}{2\pi} \exp \left( -\frac{kT \, k_v^2}{2m_p} \right)
\end{equation}
\noindent where $P_0$ is the spectrum amplitude and $N_0$ is a constant, which depends on the detector noise and the resolution.

Therefore, by fitting the predicted $P_{1,mod}$ curve to the observational data
we can determine the following parameters: 
$P_0$, \hi temperature (embedded in $\ft{f}$), and velocity parameters: 
$\alpha_v$, $L_v$ and $v_{turb}$ (all embedded in $D_z$). 
$N_0$ is estimated directly from the observational data.

We note that we omitted the influence of the instrumental channel function in Eq. (\ref{eq:FT_f}) and the regular velocity shear in Eq. (\ref{eq:P1_mod}). The instrumental channel half-width is 94 m/s, which is much less than thermal velocity of the cold phase, 670 m/s, and therefore not significant. In the direction of our observations the regular velocity shear is $b=2.34$ $m/s\cdot pc^{-1}$ which is much less than the shear resulting from turbulence, $v_{turb}/L_v = 40$ $m/s\cdot pc^{-1}$ at our largest scale, and therefore negligible.

\subsection{VCS Application} \label{sect:vcs_appl}

We calculate the velocity power spectrum $P_1$ for three different resolutions. First, we smooth the original PPV cube to resolutions of $0^\circ.125$,  $0^\circ .250$ and $0^\circ .500$. For each resolution element we calculate  $P_1$ using spatial averaging over the entire region. The resulting spectra are presented in  Fig. \ref{fig:fit} with three different types of data points.

Next we fit the observed velocity power spectra with the predicted curve. As $\alpha_\eps=3$ we can use Eq. (\ref{eq:P1_mod}) to fit the observed $P_1$ spectra with much simplified $P_{1,mod}$.

Two important issues should be noted regarding equation (\ref{eq:P1_mod}).
\begin{enumerate}
\item The warm phase of the ISM is heavily suppressed by an exponential factor in Eq. (\ref{eq:FT_f}) and its impact is negligible regardless of its abundance. Therefore, our analysis is biased to the cold neutral medium.
\item The geometric factor $g(\vc{r})$ was taken in the form of Eq. (\ref{eq:wa_clos_appr}). The inner border of the emitting layer $z_0$ was determined by the Local Bubble, while the outer border $z_1$ was calculated from the layer height, which we assumed to be 200 pc. We assumed that the cold phase is distributed evenly enough\footnote{This term needs some justification. The thermal microphysics of CNM heating and cooling mechanisms dictate that it cannot exist below a minimum pressure ${P_{min} \over k} \sim 1600$ cm$^{-3}$ K (Wolfire et al.\ 2003). With a  typical column density $N(HI) \sim 5 \times 10^{20}$ cm$^{-2}$ and  temperature $\sim 50$K, the total length along the line of sight cannot exceed $\sim 5$ pc. We assume that this thickness is statistically uniformly fragmented between $z_0$ and $z_1$.} between $z_0$ and $z_1$, with $z_0=83$ pc and $z_1=270$ pc, based on data from \citet{Lall2003}.
\end{enumerate}

The fitting is performed using Mathematica's function \verb|FindMinimum|, which implements the algorithm of steepest descent. The resulting fits are very good, as shown in Fig. \ref{fig:fit}. This process yields an estimate of the following parameters: the velocity spectral index $\alpha_v = 3.87 \pm 0.11$, the turbulent injection scale $L_v = 140 \pm 80$ pc, the VCS amplitude $P_0$, the gas temperature $T_{cold}=52 \pm 11$ K (as discussed, this is biased toward the cold medium), and the (cold phase) gas Mach number $M_{cold} = 7.7_{-0.7}^{+1.0}$.

It is remarkable, that the most of the low-$k_v$ part of the calculated spectra is in agreement with the rest of our dynamical range. We can guess that the reason is that the  correspondent $k_v$'s are mostly not affected by the warm phase. We needed to exclude only the first point, which corresponds to T=6800K and is above the estimation\footnote{When calculating this temperature we assume that for the thermal velocity projection holds $v_z=1/k_v$.}. 

Despite its negligible impact on $P_1$, the warm phase can dominate in a spectral line as whole\footnote{Zeroth and first harmonics of the high-resolution $P_1$, most likely affected by the warm phase, contain about 90\% of the total ``energy'' of $P_1$.} and in this case we can estimate its parameters too. If we know the characteristic velocity of the mean line profile $v_{total}$, we can calculate the warm phase thermal velocity as $\sqrt{v_{total}^2-v_{turb}^2}$, if we assume that $v_{turb}$ is the same for the both phases. Then we can calculate temperature and Mach number for the warm phase too. We get: the warm phase temperature $T_{warm} =  4200_{-2500}^{+1000}$K, and the warm phase Mach number $M_{warm} = 0.9_{-0.1}^{+0.5}$.

To calculate the statistical error of a fitting parameter we deviate it from its optimum and let the other parameters compensate the corresponding deviation of our target function. This gives us some other model curves, deviating from the optimal curves as well. We interpret the mean squared deviation between these two sets of model curves, normalized by variances of corresponding measured $P_1$ values, as squared normalized deviation of the parameter itself. As we know its absolute deviation, we determine its variance. By taking as a reference point the fitted model instead of the data, we separate the effect of deviation from the influence of possible systematic error and statistical scattering of data. This procedure guarantees uniqueness of the obtained solution too.

\subsection{Verification of the Fitting Procedure} \label{sect:ver}

Numerical verification of the VCS technique was presented in \citet{CL09}. However, the consistency of the fitting procedure needs to be checked as well. To do this, we can change the effective temperature of the PPV cube by convolving it over the velocity axis with a Maxwellian distribution. The temperature in the new cube is the sum of the original temperature (which we estimated to be 52 K) and that of the Maxwellian distribution. Temperature is the only fitting parameter we can easily change.

We have convolved our data with the $104$ K Maxwellian distribution. Our fitting procedure gave $T=154$ K, $\alpha_v = 3.87$, $L_v=137$ pc, $v_{turb}=5.14$ km/s (the original parameters were $T = 52$ K, $\alpha_v = 3.87$, $L_v=138$ pc, $v_{turb}=5.15$ km/s). The expected temperature is $52+104=156$ K, very near to the value the procedure retrieved. Therefore, we see a good agreement between the estimate from our fitting procedure and the theoretically expected value when we increased the temperature by a factor of 3. All other parameters remained very near to their original values, which demonstrates good stability of the solution. 

\subsection{Verification of the Derived Temperature} \label{sect:ver_temp}

Figure \ref{fig:temps} exhibits the derived HI spin temperature $T_x$ versus $VLSR$ for 
6 NVSS continuum sources that lie in the analyzed region; 4 sources have 
two measurable velocity components, providing the total of 10 samples 
shown. The sources all have 1.4 GHz flux densities exceeding 0.75 Jy and 
constitute the complete set of sources in this region that yield 
reliably-detected 21-cm absorption line profiles.

Each datum consists of two by-eye estimates, one a lower limit and the 
other an upper limit. The upper limit assumes that all the emission at the 
peak of the absorption line (the ``expected emission temperature'' 
$T_{B,exp}$) arises from gas at a uniform spin temperature. This situation 
is generally unrealistic, however, because unrelated gas along the line of 
sight contributes to $T_{B,exp}$; hence, this assumption provides us with 
the upper limit. The lower limit arises by decomposing the emission 
profile into Gaussian components, and assigning to $T_{B,exp}$ only the 
emission associated with the component that represents the absorbing gas. 
This implicitly assumes that the unrelated emission from warmer gas is not 
absorbed by the cold gas, so the warm gas lies in front of the cold 
absorbing gas; because some warm gas might lie behind, this assumption 
provides us with the lower limit. The limits in Figure \ref{fig:temps} are approximate 
because they were determined by eye, not by least-squares fitting of 
Gaussian components, but the errors in these by-eye estimates are 
considerably smaller than the ranges of temperature shown.

The data in Figure \ref{fig:temps} refer to cold gas. Typically, a significant fraction 
of the gas is warm and produces no detectable 21-cm line absorption. This 
warm gas is not represented in Figure \ref{fig:temps}. The limit of detectability 
depends on the flux of the continuum background source, and if this region 
had stronger sources we would have been able to plot data with warmer 
temperatures than Figure  \ref{fig:temps}.

The bias towards cold gas that is introduced by these measurement 
considerations---which are observationally based---is similar in spirit to 
the bias inherent in our statistical analysis, which is also more 
sensitive to cold gas than warm. Thus, a comparison of the data in Figure 
 \ref{fig:temps} with our statistically-derived spin temperature of 52 K is meaningful. 
In Figure  \ref{fig:temps}, half of the data are consistent with the 
theoretically-derived 52 K; the other half are warmer. Given the 
similarity between the observational and theoretical biases, we regard 
this agreement as satisfactory.

\section{Discussion and Summary} \label{sect:dis}

\subsection{Retrieved Parameters} \label{sect:pars}

Applying the VCS technique to the $6.5^\circ \times 6.5^\circ$ site centered at $l=151^\circ$, $b=-49^\circ$ we have determined the following \hi parameters: 
\begin{itemize}
 \item velocity spectral index $\alpha_v = 3.87 \pm 0.11$
 \item density spectral index $\alpha_{\eps} = 3.0 \pm 0.1$
 \item injection scale $L_v = 140 \pm 80$ pc
 \item cold phase temperature $T_{cold} = 52 \pm 11$ K
 \item cold phase Mach number $M_{cold} = 7.7_{-0.7}^{+1.0}$
\end{itemize}
In addition, under the assumption that the warm medium dominates the line at low-$k_v$'s, we get the warm phase temperature $T_{warm} = 4200_{-2500}^{+1000}$ K and the warm phase Mach number $M_{warm} = 0.9_{-0.1}^{+0.5}$.

The derived $T_{cold}$ is very similar to what is typically assumed and measured for the CNM. For example, Heiles \& Troland (2003) used an HI absorption/emission survey of 79 sources and found that the CNM spin temperature histogram peaks at about 40 K, while its median, weighted by column density, is 70 K. It is also in satisfactory agreement with our measurements of spin temperature within the analyzed region (see Sect \ref{sect:ver_temp}). The inferred $T_{warm}$ is in agreement with the typical temperature of the WNM of $\sim5000$ (Wolfire et al. 2003). Similarly, $T_{warm}$ in the range $500-5000$ K was estimated observationally by Heiles \& Troland (2003) to occupy 48\% of the WNM and corresponds to the thermally-unstable warm gas. Heiles \& Troland (2003) also found that $M_{cold} \approx 3$ for the Milky Way, but has a large scatter and ranges from $\sim1$ to $\sim7$. Our estimated $M_{cold}$ is at the boundary of the Heiles \& Troland (2003) range.

Under the assumption that the cold gas has a uniform distribution from 83 pc to 270 pc along the line of sight, we estimated the injection scale of turbulence $L_v = 140 \pm 80$ pc. This is in agreement with the expected value of 100 pc associated with supernova explosions (compare to \citet{Hav08}).

It is interesting that our velocity and density spectral indices are different from indices derived for the Ursa Major high-latitude cloud, $\alpha_v=\alpha_{\epsilon}=3.6\pm0.2$ by Miville-Deschenes et al. (2003).These authors used a very different approach, velocity centroids and the assumption that density and velocity fields are Gaussian. While in the case of Ursa Major field velocity and density indices are similar, in our case there are very different: $\alpha_v \approx 3.9$ and $\alpha_{\eps} \approx 3$. In addition, studies of centroids (Lazarian \& Esquivel 2003, Esquivel \& Lazarian 2005, Ossenkopf et al. 2006, Esquivel et al. 2007) showed that ordinary centroids used in Miville-Desch{\^e}nes et al. 2003 may represent only velocity statistics in subsonic turbulence. This is unlikely for most of \hi, which has an admixture of cold and warm gas.

Generally, the estimated parameters are reasonable and this suggests that the VCS technique could be used to estimate gas temperature and turbulent properties directly from \hi emission profiles, instead of obtaining \hi absorption spectra. However, we worked here with only a single region and further testing with observational data is essential to check VCS's reliability.

\subsection{Advantages and Limitations of our Approach}

While the analysis of fluctuations in channel maps has been a relatively standard technique, the analysis of the fluctuations within PPV cubes along the velocity direction is quite a new approach. Naturally, we faced many new problems in this situation to which we have to present our solutions.

The most fundamental problem that we face dealing with the VCS technique is the necessarily limited inertial interval. We can demonstrate this assuming that the turbulence is Kolmogorov. In this case $v\sim l^{1/3}$ and an inertial range of $10^3$ in terms of $l$ translates to just one decade of inertial range in the velocity domain. While the actual astrophysical turbulence spans over many decades (see Armstrong et al. 1994), the measurements of the turbulent velocity fluctuations become very challenging because of the thermal broadening of lines. The latter depends on the mass of the species and is most prominent for atomic hydrogen. This was the reason that we found asymptotic studies not useful and adopted the fitting procedure described in the paper. We expect modifications of this procedure will be used in the future with other species.

However, in the no-asymptotic case we needed to fit several parameters. The pros and cons of the fitting procedure are interrelated. In general, it may be considered safer to measure just the index of the power slope, as it is prescribed in the VCA technique (see Stanimirovic \& Lazarian 2001) rather than to fit several parameters simultaneously, including the injection scale and the gas temperature. However, the Doppler-shifted lines contain all this information and a successful fitting procedure can provide more than just the velocity spectral slope in question. In fact, we are developing a similar fitting procedure for the VCA technique (Chepurnov \& Lazarian, in preparation), which shows advantages compared to the usual employment of the channel map data.

An additional advantage of the fitting procedure is that some parameters of the model (e.g. temperature, injection scale, turbulent velocity) can be independently studied and tested. At the same time, in the situations when these parameters are not available by other means, the fitting procedure can provide estimates of them as discussed in LP06. Our verification procedure in Sect. \ref{sect:ver} is encouraging in this respect.

The fact that the information for the VCS is taken from the fluctuations along V-direction of the PPV cubes allows us studying spatially localized regions of turbulence and map the distribution of turbulence within the studied turbulent volume, which is another advantage of VCS. In this respect, VCS can be successfully used in conjunction with other statistical tools which allow to get insight into turbulence (see Burkhart et al. 2009).

For our analysis we have chosen high latitude Galactic \hi. This gives advantages both through high resolution, which allows studies of different resolution limits of the VCS technique, and it allows us to worry less about the effects of confusion that arise when \hi in the plane of the Milky Way is studied. However, the downside of this is the necessity to deal with a more complex observational geometry, where fluctuations of different physical size are seen at the same angular scale (i.e. perspective). We have formalized the influence of observational geometry by introducing a geometric term, which contributes to the integrand in the expression for $P_1$.

The column density map shows spectral index of 3 with very good confidence. On one hand, this allows us to omit density factor in calculations of $P_1$, if we take this value as a true estimation of the density spectral index. On the other hand, the VCA analysis for the absorbing medium predicts exactly this behavior for the optically thick case for any density spectral index (see LP04 for details). 

At the same time the temperature measured by the telescope is low enough to formally preclude the case of self absorption. Nevertheless, it may still happen that \hi is not self-absorbing in terms of total emission and is self-absorbing in terms of fluctuations of PPV statistics. For instance, it is well established (see LP00, LP06) that the PPV fluctuations are dominated by cold gas, while the contribution of the fluctuations in warm gas is exponentially suppressed. Thus, if the cold gas, which for high latitude sampled by our observations constitutes a small fraction of the total mass, still dominates the PPV fluctuations, we may have the situation that we observe. In this case the density factor is undetermined. We can assume that the density spectrum is steep: in this case the density term can be factored out too (for asymptotic studies). However, the possibility of shallow density, i.e. $\alpha_{\eps}<3$, exists. In this case the spectral index of velocity is not $\alpha_v\approx 3.9$, but $\alpha_v=3+0.9(\alpha_\eps -2)$. The bracket in this case is less than $1$, which means that the spectral index of velocity is $\alpha_v<3.9$. 

The Kolmogorov index corresponds to $\alpha_v\approx 3.7$, but turbulence with this index is known to produce the spectral index of density $\alpha_\eps \approx 3.7$ (Cho \& Lazarian 2003). One requires supersonic turbulence to get $\alpha_{\eps}<3.7$ (Beresnyak et al. 2005). Such turbulence corresponds to $\alpha_v>3.7$. To satisfy these requirements one should have $2.77<\alpha_\eps<3$, which provides constraints on the density spectrum in the vicinity of the $\alpha_\eps=3$ that we assumed.  

On the basis of the above, we assumed that the fluctuations we measure are due to the velocity fluctuations only and correspond to the spectral index $\alpha_v\approx 3.9$, which is steeper than the Kolmogorov index. As our analysis is exponentially biased towards cold gas the spectrum measured is mostly of turbulence in the cold gas. In the situation when a turbulence in cold gas is a part of a large scale cascade, as it is generally assumed, the turbulence in the cold gas is supersonic and the formation of the shock-type velocity spectrum observed is not surprising at all. 

While our paper were in the process of refereeing, we learned about the new paper by Padoan et al. (2009) submitted to ApJ. The paper also uses VCS technique, but it does not provide the fit to the data as we do in this paper. Instead, Padoan et al. (2009) compare the VCS power slope to the asymptotic predictions in Lazarian \& Pogosyan (2000). We feel that this way of obtaining the turbulence power spectrum is subject to larger errors compared to the technique we use in the paper. Indeed, the dynamical range of the VCS fluctuations is rather restricted which limits the applicability of fitting of the asymptotic slope. 

However, the spectral index value, calculated in assumption of asymptotic regime can be used as a lower limit for $\alpha_v$, if the emissivity term is negligible. Making a stronger statement needs a posteriori check by direct calculation of $P_1$. For instance, performing the same procedure as in Padoan et al. (2009) we would get the spectral index of turbulence of $3.81\pm 0.04$, which deviates by a systematic error from the numbers $3.87 \pm 0.11$ we obtain using our approach (see Appendix for the details).

\subsection{Summary}

We have applied the new VCS technique to the Arecibo high latitude data and obtained the spectrum of velocity, which is steeper than the Kolmogorov one. The steeper turbulent velocity spectrum indicates the importance of shocks in the media, which are expected to make the spectrum of density shallower than the Kolmogorov density. This is the effect that we register studying the turbulent density. Our application of the VCS technique uses model fitting procedure, which allows us to evaluate the injection scale of the turbulence, the temperature of the cold media, turbulent velocity and Mach number. Assuming that the warm medium dominates the line at low $k_v$'s, we estimated the temperature and Mach number of the warm phase too. The obtained parameters for the region of the study are given in \S \ref{sect:pars}.

\acknowledgments{AC and AL acknowledge the support of the NSF grant AST0808118 and the Center for Magnetic Self-Organization in Laboratory and Astrophysical Plasmas. SS acknowledges support by the NSF grant AST-0707679.}


\appendix

\section{Applicability of Asymptotics} \label{sect:asymp}

LP00 and LP09 presented their final results in terms of asymptotics of $P_1$ for large $k_v$. While this is advantageous from theoretical point of view, it presents some problems related to the analysis of observational data. Below we show that the use of asymptotics may require higher resolution and larger dynamical range than it is available from observations.    

Let us check if the asymptotical approach is possible for our data. To do that we can calculate velocity spectral index, as if the asymptotics is applicable, and compare it to the value obtained by our fitting procedure. The correspondent linear regression is shown on Fig. \ref{fig:p1_hr_fit}. We can see, that there is some disagreement between the obtained spectral index $\alpha_v=3.81\pm 0.04$ and the value, obtained by the model fitting ($\alpha_v=3.87\pm 0.11$). I.e. we can conclude that in our case the calculation assuming high-resolution asymptotic regime produces a systematic error bigger than the statistical error of the asymptotic $\alpha_v$.

\begin{figure}
\begin{center}
\plotone{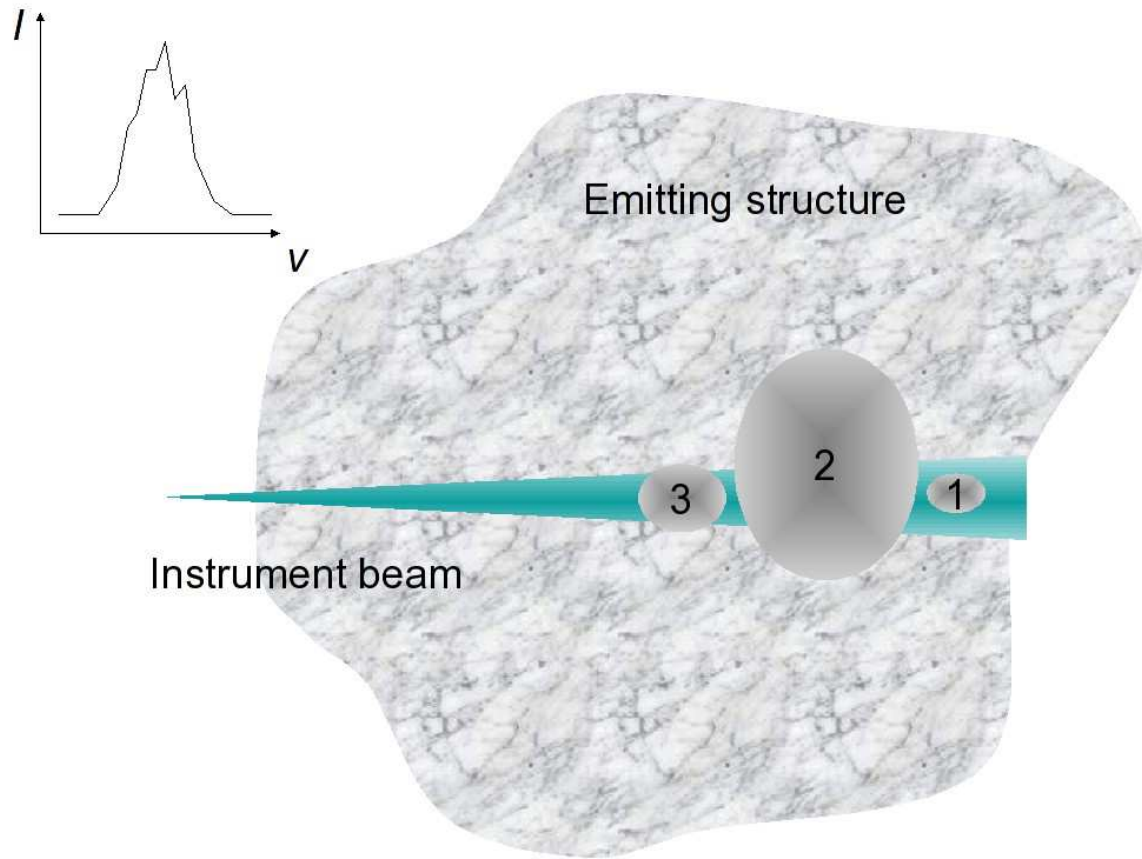}
\end{center}
\caption{VCS technique: effects of resolution. The fluctuations {\it along} the velocity coordinate are analyzed. Eddies within the telescope size beam, e.g. eddy  1, are in a low resolution mode. Eddies with the size exceeding the one of the beam, e.g. eddy 2, are in the high resolution mode. The conical geometry of the emitting volume within the beam affects our analysis too.\label{fig:geo}} 
\end{figure}

\begin{figure}
\begin{center} 
\plotone{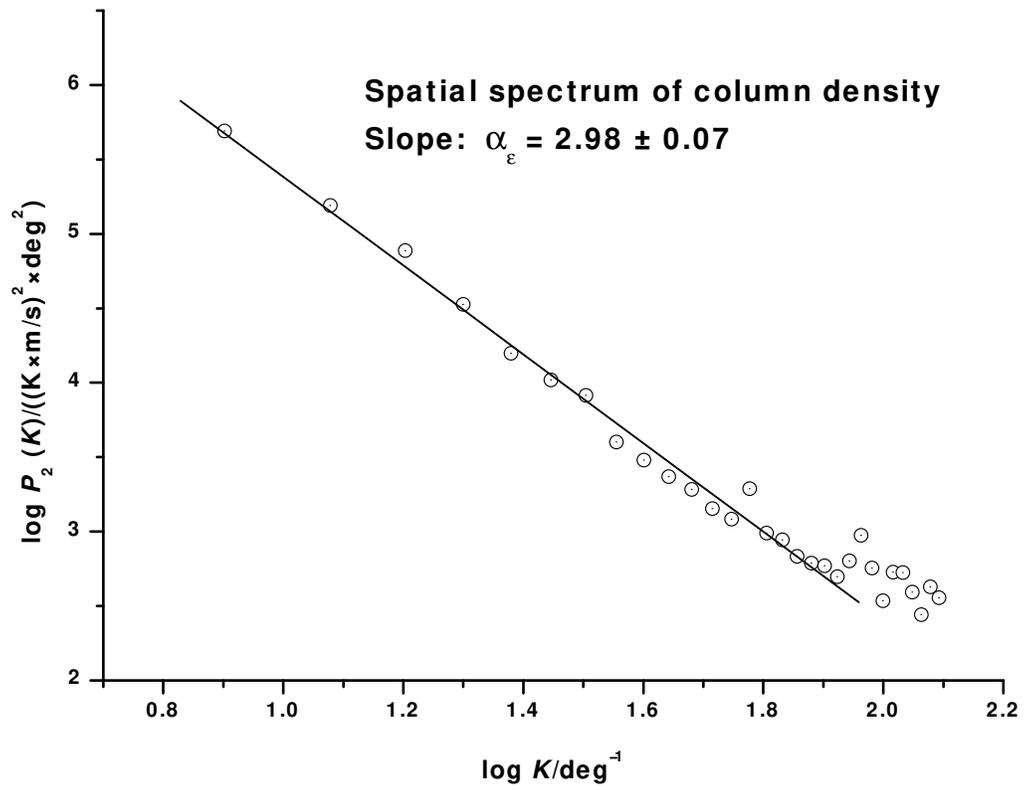}
\end{center}
\caption{Spatial power spectrum of the \hi column density image. The estimated spectral index of 3 results in a weak dependence of the 3D density correlation function on radius-vector.\label{fig:rho}}
\end{figure}

\begin{figure}
\begin{center} 
\plotone{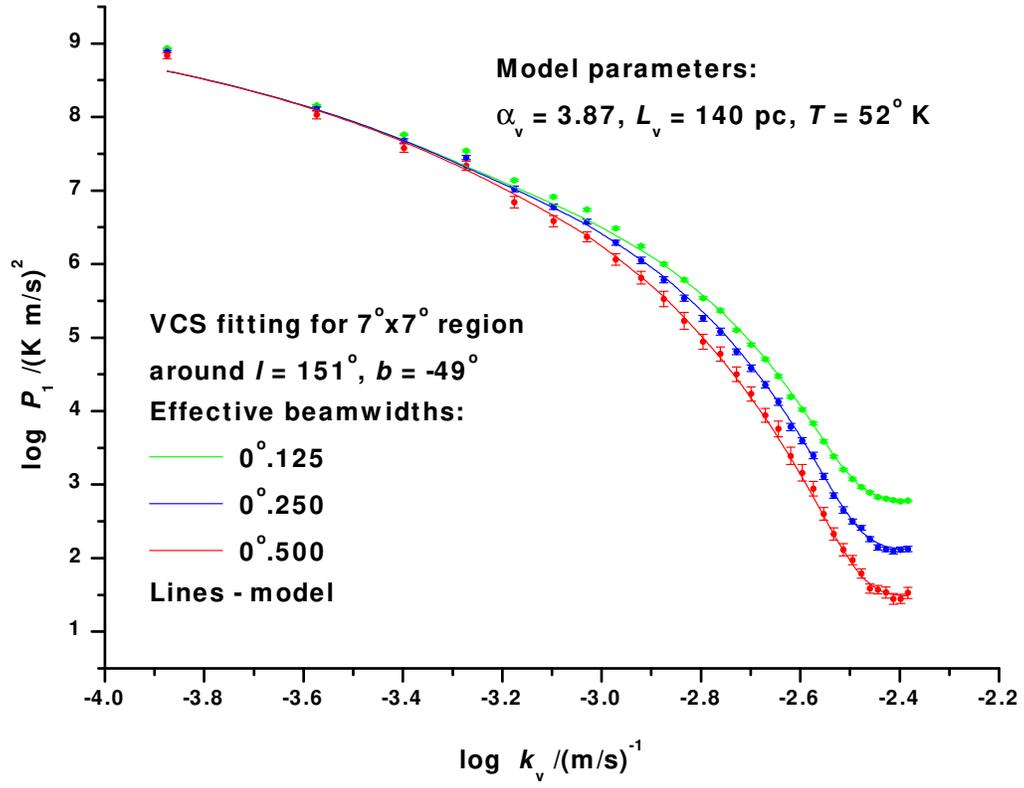}
\end{center}
\caption{Fitting of the model VCS spectra to $P_1$ data, obtained from GALFA-\hi data cube for different effective resolutions. The leftmost points, most likely affected by the warm phase (their abscissa corresponds to the kinetic temperature of 6800K) are excluded from the fitting. \label{fig:fit}}
\end{figure}

\begin{figure}
\begin{center} 
\plotone{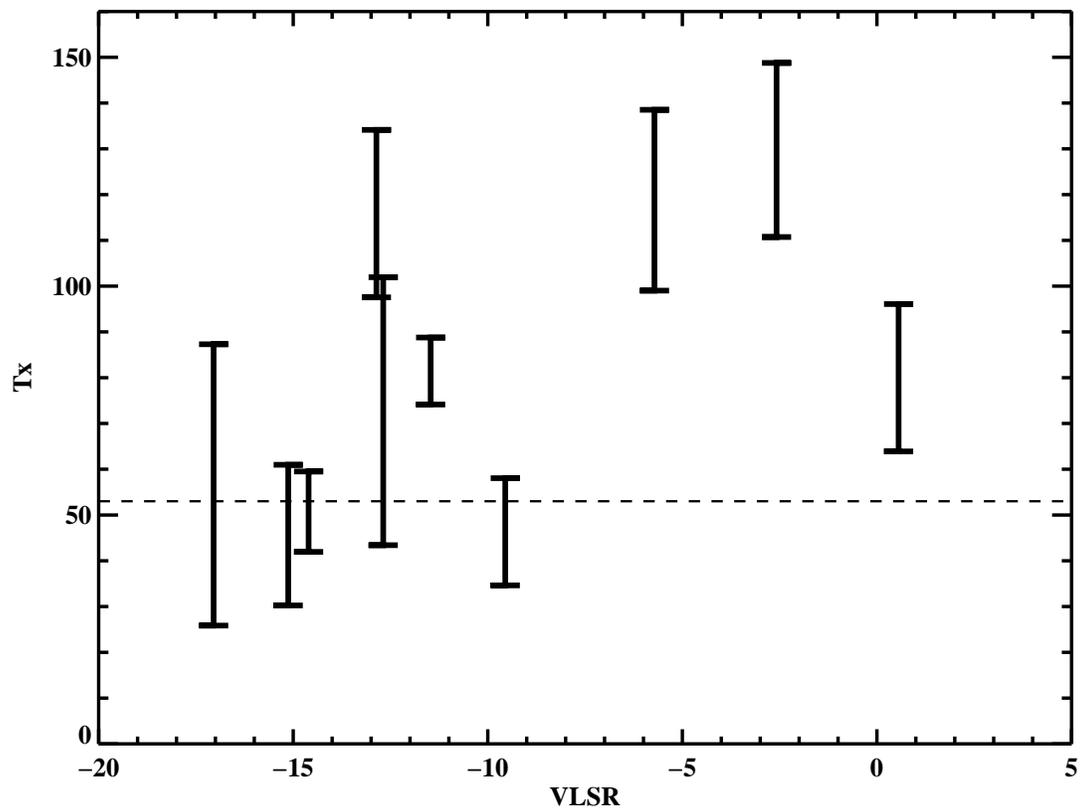}
\end{center}
\caption{This figure exhibits the derived \hi spin temperature $T_x$ versus $VLSR$ for 6 NVSS continuum sources that lie in the analyzed region; 4 sources have two measurable velocity components, providing the total of 10 samples shown. The dashed line is the temperature derived from our statistical analysis. See Tab. \ref{tab:temps} for the correspondent estimation data. \label{fig:temps}} 
\end{figure}

\begin{figure}
\begin{center} 
\plotone{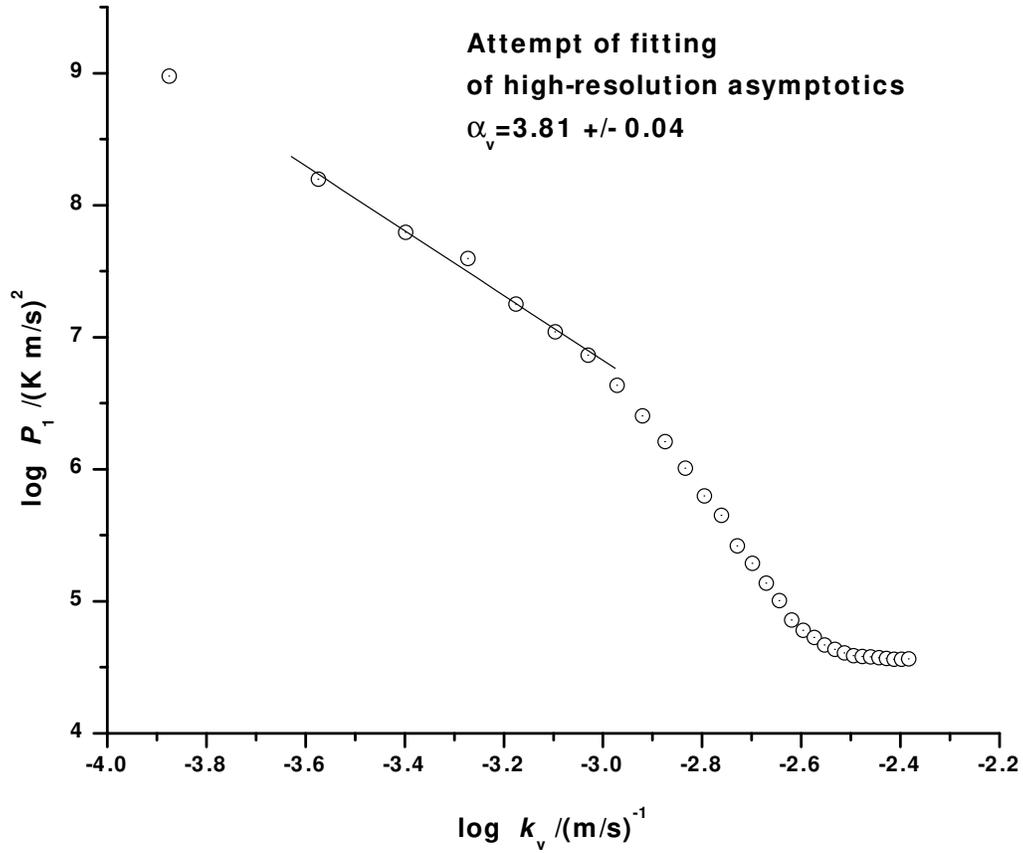}
\end{center}
\caption{Attempt of fitting of high-resolution asymptotics. The derived velocity spectral index $\alpha_v=3.81\pm 0.04$ deviates by a systematic error from the spectral index obtained by the model fitting ($\alpha_v=3.87\pm 0.11$). The obtained value shows some systematic error and is actually a lower limit for $\alpha_v$ (see Fig. \ref{fig:p1_ratio}, left panel) \label{fig:p1_hr_fit}}.
\end{figure}

\begin{figure}
\begin{center}
\plottwo{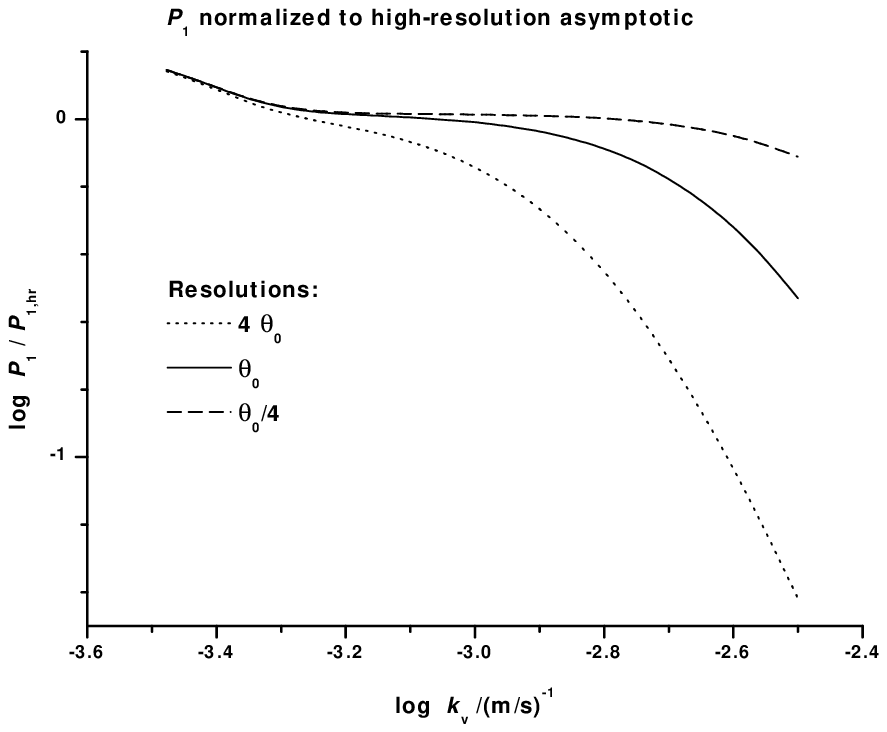}{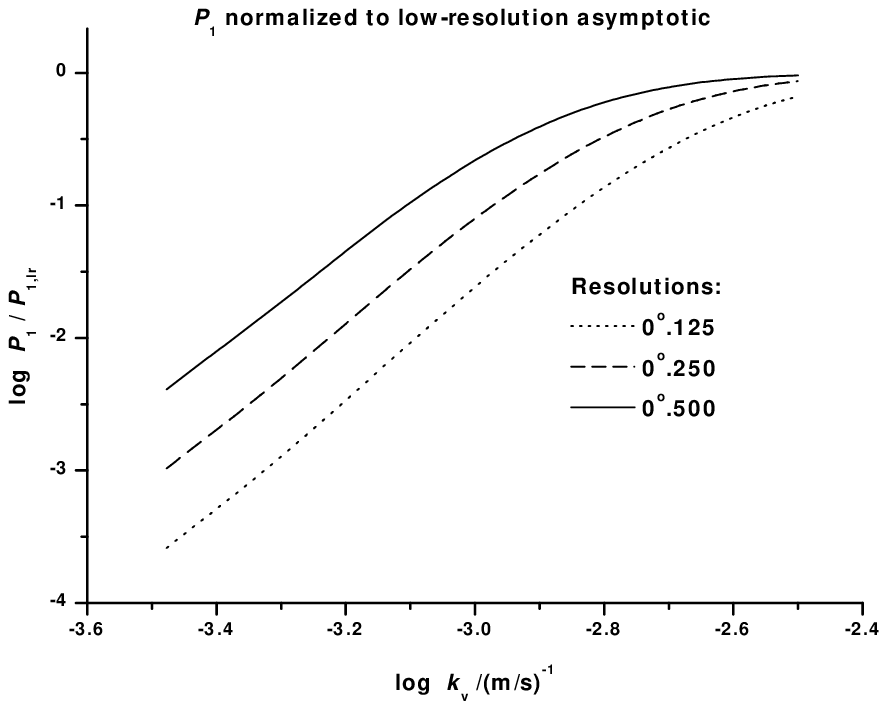} 
\end{center}
\caption{Left: $P_1$ normalized to the high-resolution asymptotic for the instrumental resolution (solid line) and resolutions 4 times lower (dotted line) and 4 times higher (dashed line). The assumption of the high-resolution asymptotic regime is applicable for the latter case only, for other cases there is some systematic error, more significant for the lower resolution. Right: $P_1$ normalized to the low-resolution asymptotic for the resolutions $0^o.125$ (dotted line) $0^o.250$ (dashed line) and $0^o.500$ (solid line). The assumption of the low-resolution asymptotic regime is not applicable. \label{fig:p1_ratio}}
\end{figure}

As it is clear from Figure \ref{fig:geo}, the notion of the high and lower resolution changes as the eddies of different sizes are studied. For the largest eddies and therefore for the small $k_v$ we are in the limit of high resolution. However, this changes as we go to larger $k_v$. We can also compare the model $P_1$ with the predicted asymptotic for different resolutions, see Fig. \ref{fig:p1_ratio}, left panel. We could see there, that if the resolution were 4 times higher than the one of our instrument, it were possible to compute velocity spectral index directly from the $P_1$ slope. 

In what regime we are may not be obvious from the very beginning. If to find the velocity statistics we use the high-resolution asymptotics it is advisable to have a posteriori check if the assumption of the high resolution is applicable. This complicates the analysis. On the contrary, the fitting of a model $P_1$, as it is done in the present paper, although it is being harder to implement, is self-sufficient and more reliable. However, as it can be seen from Fig. \ref{fig:p1_ratio}, left panel, asymptotic calculation can be used to get a lower limit for $\alpha_v$, if we neglect the emissivity term.

What we said above is related to the high-resolution asymptotic solution. What about low-resolution asymptotics obtained in LV00 and LV06? We can also check the low-resolution asymptotics (see Fig. \ref{fig:p1_ratio}, right panel). We observe that such regime is not present for the degraded resolutions we used for the model fitting. Theoretically it is possible to degrade resolution until the slope of $P_1$ saturates, and then calculate the velocity spectral index using the low-resolution asymptotics. However, in our case the high-$k_v$ part of $P_1$ is affected by the temperature term, and this approach becomes unreliable for our data. In addition, $P_1$ gets noisier if we degrade the resolution.

All in all, the asymptotic formulae may not be straightforward to use with the real observational data. Therefore, we advocate the numerical procedure presented in the paper as a more reliable way of handling our data set. This does not mean that the asymptotical analytical solutions are not useful. First of all, they provide a proper insight into qualitative properties of the PPV data cubes. In addition, if one uses a spectral line of a heavier species and have enough statistics and dynamical range for calculation of the low-resolution $P_1$, the asymptotical approach is applicable and self-sufficient. In general, we do not want to confront the asymptotic solutions and the formulae that we use. One should remember that the asymptotic solutions were obtained by the evaluating for high $k_v$ the integral expressions employed in this work.


\newpage

\begin{deluxetable}{llllll}
\tablecaption{Spin temperature estimation data for Fig. \ref{fig:temps} \label{tab:temps}}
\tablehead{
  \colhead{ } & \colhead{$l$} & \colhead{$b$} & \colhead{VLSR, km/s} & \colhead{$T_{min}, K$} & \colhead{$T_{max}, K$}
}
\startdata
       0   &   149.684  &   -47.4998  &   -9.55142  &    34.6271  &    58.0417 \\
       1   &   151.554  &   -49.6997  &   -12.6884  &    43.4076  &    101.944 \\
       2   &   151.554  &   -49.6997  &   -5.71733  &    99.0173  &    138.530 \\
       3   &   153.901  &   -51.2807  &   -17.0453  &    25.8466  &    87.3100 \\
       4   &   153.901  &   -51.2807  &   -12.8627  &    97.5539  &    134.139 \\
       5   &   155.258  &   -45.8268  &   -14.6054  &    41.9442  &    59.5051 \\
       6   &   155.258  &   -45.8268  &   -2.58035  &    110.725  &    148.773 \\
       7   &   157.784  &   -48.1993  &   -11.4685  &    74.1393  &    88.7734 \\
       8   &   157.784  &   -48.1993  &   0.556630  &    63.8954  &    96.0905 \\
       9   &   159.721  &   -49.1208  &   -15.1283  &    30.2368  &    60.9685
\enddata
\tablecomments{
In cases of more than one line for the same position, there are two recognizable velocity components.
}
\end{deluxetable}

\end{document}